\documentclass[11pt]{article}
\newtheorem{theorem}{Theorem}
\newtheorem{proposition}{Proposition}
\newtheorem{lemma}{Lemma}
\newtheorem{corollary}{Corollary}
\newtheorem{definition}{Definition}

\newcommand{\begsection}[1]{\setcounter{equation}{0}\section{#1}}

\def\C{{\mathcal C}}

\def\R{{\mathcal R}}
\def\Z{{\mathcal Z}}
\def\T{{\mathcal T}}

\def\quadratino{\hfill
\vbox{\hrule\hbox{\vrule\vbox to 7 pt {\vfill\hbox to 7 pt
{\hfill\hfill}
\vfill}\vrule}\hrule}\par}

\def\Sc{Schr\"o\-din\-ger}
\def\hp{{\hbar}}

\def\la{\langle}
\def\be{\begin{equation}}
\def\ee{\end{equation}}
\def\ra{\rangle}
\def\ds{\displaystyle}

\def\om{\omega}
\def\Om{\Omega}
\def\ep{\epsilon}

\def\quadratino{
\hfill\vbox{\hrule\hbox{\vrule\vbox to 7 pt {\vfill\hbox to
7 pt {\hfill\hfill}\vfill}\vrule}\hrule}\par}

\def\F{{\mathcal F}_{\rho,\sigma}}

\def\quadratino{
\hfill\vbox{\hrule\hbox{\vrule\vbox to 7 pt {\vfill\hbox to
7 pt {\hfill\hfill}\vfill}\vrule}\hrule}\par}

\def\sleq{\leq\kern-4pt\cdot}
\def\sgeq{\null\kern+1pt\cdot\kern-4.5pt\geq}
\def\om{\omega}
\def\be{\begin{equation}}
\def\ee{\end{equation}}

\begin{document}
\baselineskip=19pt
\begin{center}
{\large\bf A LOCAL  QUANTUM VERSION OF THE \\
KOLMOGOROV THEOREM}
\end{center}
\vskip 13pt
\begin{center}
David Borthwick\footnote{Supported in part by NSF grant DMS-0204985.
Department of Mathematics and Computer Science,
Emory University, Atlanta 30322 (U.S.A.). (davidb@mathcs.emory.edu)},
Sandro Graffi\footnote{
Dipartimento di Matematica,  Universit\`{a} di Bologna, 40127 Bologna
(Italy).
 (graffi@dm.unibo.it)} 
\end{center}
\begin{abstract}
\noindent
Consider in $L^2 (\R^l)$ the  operator family
$H(\epsilon):=P_0(\hbar,\omega)+\epsilon Q_0$.  
$P_0$ is the quantum harmonic oscillator with diophantine frequency vector
$\om$, $Q_0$ a  bounded 
pseudodifferential operator with symbol holomorphic  and
decreasing to  zero at infinity, and $\ep\in\R$.  Then there exists 
$\ep^\ast >0$  with the property that if $|\ep|<\ep^\ast$ there
is  a diophantine 
frequency $\om(\ep)$ such that  all eigenvalues $E_n(\hbar,\ep)$ of
$H(\ep)$ near $0$  are
given by the quantization formula 
$E_\alpha(\hbar,\ep)={\cal E}(\hbar,\ep)+\la\om(\ep),\alpha\ra\hbar
+|\om(\ep)|\hbar/2 + \ep O(\alpha\hbar)^2$, where
$\alpha$ is an $l$-multi-index.
\end{abstract}
\vskip 1cm   
%
 
%


\begsection{Introduction and statement of the results}
\setcounter{equation}{0}%
\setcounter{theorem}{0}%
\setcounter{corollary}{0}%
Denote by
${\cal F}_{\rho,\sigma}$ the set of all   functions
$f(x,\xi):\R^{2l}\to\C$ with finite $\|f\|_{\rho,\sigma}$ norm for some
$\rho>0$, $\sigma >0$ (see Section 2 for the definition and examples).
Any $f\in {\cal F}_{\rho,\sigma}$ is analytic on $\R^{2l}$ and extends to a
complex analytic function in the region $|\Im{z_i}|\leq a_i|\Re {z_i}|$ for
suitable $a_i>0$; moreover $|f(z)|\to 0$ as $|z|\to+\infty$. Here
$z:=(x,\xi)$. 

Let $\Phi_{\rho,\sigma}$ denote the class of semiclassical
Weyl pseudodifferential operators $F$ in $L^2(\R^l)$ with symbol $f(x,\xi)$
in
${\cal F}_{\rho,\sigma}$; 
namely, (notation as in \cite{Ro}) 
\begin{eqnarray}
\label{Weyl}
(Fu)(x)&:=&Op^W_h(f(x,\xi))u(x)
\\
\nonumber
&=&\frac{1}{h^l}\int\!\!\!\int_{\R^l\times\R^l}e^{i\la
(x-y),\xi\ra/\hbar} f((x+y)/2,\xi)u(y)\,dyd\xi,\; u\in{\cal S}(\R^l).
\end{eqnarray}
It follows directly from the definition of $\|f\|_{\rho,\sigma}$
in (\ref{norma}) that $F\in \Phi_{\rho,\sigma}$ extends to a conti\-nuous 
operator in $L^2(\R^l)$, with
\be
\|F\|_{L^2\to L^2}\leq \|f\|_{\rho,\sigma}.
\ee
Consider in $L^2(\R^l)$ the
operator family
$H(\ep)=P_0(\hbar,\om)+\ep Q_0$ and assume:
\begin{itemize}
\item[(A1)]  $P_0(\hbar,\om)$ is the harmonic-oscillator \Sc\ operator  
 with frequencies  $\om\in [0,1]^l$:
\be
\label{HO}
P_0(\hbar,\om)u=-\frac12\hbar^2\Delta u+[\om_1^2x_1^2+\ldots+\om_l^2x_1^2]u,
\;\; D(P_0)=H^2(\R^l)\cap L_2^2(\R^l).
\ee
\item[(A2)] $Q_0\in \Phi_{\rho,\sigma}$; its  symbol 
$q_0(x,\xi)=q_0(z)$ is real-valued for $z=(x,\xi)\in\R^\times\R^l$, 
and 
$q_0(z)=O(z^2)$ as $z\to 0$.
\item[(A3)]  There exist $ \tau>l-1,\gamma >0$ such that
\begin{equation}
\label{Diofanto}
\la{ \om}, { k}\ra \geq {\gamma}{|{k}
|^{-\tau}}, \quad \forall {
k}\in 
\Z^l\setminus\{0\}, \quad |k|:=|k_1|+\ldots+|k_l|,\;
\om:=(\om_1,\ldots,\om_l).
\end{equation}
Denote $\Omega_0$ the set of all $\om\in[0,1]^l$ fulfilling
(\ref{Diofanto}), and $|\Omega_0|$ its measure. It is well known that
$|\Omega_0|=1$.
\end{itemize}
Under the above assumptions the operator family $H(\ep)$  defined on $D(P_0)$ 
is 
self-adjoint with pure-point  spectrum $\forall\,\ep\in\R$: 
${\rm Spec}\,(H(\ep))={\rm Spec}_p\,(H(\ep))$.  Moreover 
(\ref{Diofanto})
entails  in  particular the rational independence of the components of
$\om$ and hence the  simplicity of ${\rm Spec}(P_0)$ and its  
density  in $\overline{\R}_+:=\R_+\cup\{0\}$. Clearly,
$P_0$ is a  semiclassical pseudodifferential operator of order $2$ with
symbol 
\begin{eqnarray}
\label{azioni}
p_0(x,\xi)=\frac12(|\xi|^2+|\om x|^2)
=\frac12\sum_{k=1}^l\om_kI_k(x,\xi), 
\;
I_k(x,\xi):=\frac{1}{2\om_k}[\xi_k^2+\om^2_kx_k^2], \; k=1,\ldots,.
\end{eqnarray}
\vskip 0.3cm\noindent
\begin{theorem}
\label{mainth}
Let (A1-A3) be verified; let $h^\ast>0$.  Then 
given $\eta>0$ there exist $\ep^\ast >0$ and, for all $\ep\in 
[-\ep^\ast,\ep^\ast]$,   $\Omega^\ep 
\subset\Omega_0$
independent of $(\hbar\in[0,\hbar^\ast]$, $\eta)$
and $\om(\hbar,\ep)\in\Omega^\ep$, such  that if 
$|\alpha\hbar|<\eta$ the spectrum of 
$H(\ep)$  is given by the quantization formula
\be
\label{quantiz}
E_\alpha(\hbar,\ep)={\cal E}(\hbar;\ep)+\la
\om(\hbar,\ep),\alpha\ra\hbar+\frac12|\om(\hbar,\ep)|\hbar+\ep 
{\cal  
R}(\alpha\hbar,\hbar;\ep).
\ee
Here:
\par\noindent
1. ${\cal E}(x;\ep):[0,h^\ast]\times  
[-\ep^\ast,\ep^\ast]\to\R$ is 
continuous in 
$x$ and analytic in $\ep$, with ${\cal E}(x,0)=0$, 
${\cal E}(0;\ep)=0$;
\par\noindent
2.  $\om(x;\ep): 
[0,h^\ast]\times [-\ep^\ast,\ep^\ast]\to\R$ is continuous in
$x$ and  analytic in 
$\ep$ 
with
$\om(x;0) = \om$.
\par\noindent
3. ${\cal R}(x,y,\ep): \overline{\R}_+^l\times 
[0,h^\ast]\times
[-\ep^\ast,\ep^\ast]\to\R$ is  
continuous in $(x,y;\ep)$ and 
such that
\be
\label{resto}
|{\cal R} (x,y;\ep)|=O(|x|^2),
\ee
uniformly with respect to $(y,\epsilon)$. 
\par\noindent
4.  $|\Omega^\ep-\Omega_0|\to
0$ as
$\ep\to 0$.
\end{theorem}
\par\noindent
The uniformity in $\hbar$ of the estimates needed to prove Theorem 1.1 yields 
 in this particular 
setting  a formulation of  
Kolmogorov's theorem equivalent to  that of  \cite{BGGS}:
\begin{corollary}
\label{mainc}
Let  $\ep^\ast$, $\Omega^\ep$, 
${\cal E}(x;\ep)$, $\om(x;\ep)$  
be as above. Then  $\forall\,\ep$  there is  an analytic  
canonical transformation 
$(x,\xi)=\psi_{\ep} (I,\phi)$ of $\R^{2l}$ onto $\overline{\R}_+^l\times\T^l$ 
such that
\be
\label{classico}
(p_\ep \circ\psi)(I,\phi)={\cal E}(\ep)+\langle 
\om(\ep),I
\rangle +\ep 
\tilde{\cal R}(I,\phi;\ep)
\ee
Here ${\cal E}(\ep):={\cal E}(0;\ep)$, 
$\om(\ep):=\om(0;\ep)\in\Om^\ep$; 
 $\tilde{\cal R}(I,\phi;\ep)=O(I^2)$ as $I\to 0$ 
uniformly in $\phi$.
\end{corollary}
{\bf Remarks}
\begin{enumerate}
\item The form (\ref{classico}) of the Hamiltonian entails that a quasi 
periodic-motion with diophantine perturbed frequency 
$\om(\ep)\in\Om^\ep$ exists on the 
perturbed torus $I=0$;  equivalently, a  quasi periodic motion with
frequency $\omega(\ep)\in\Omega^\ep$ exists on the unperturbed torus with 
parametric 
equations 
$(x,\xi)=\psi_{\ep}(0,\phi)$. 
 Making 
$I=\alpha\hbar$ (\ref{quantiz}) represents the 
quantization of 
the r.h.s. of (\ref{classico}). 
In 
the formulation of \cite{BGGS} a  quasi periodic motion with the 
unperturbed
frequency $\omega\in\Omega$ exists on an unperturbed torus with parametric 
equations
$(x,\xi)=\psi_{\ep}(0,\phi)$.  The selection of the diophantine frequency 
within $\Omega$ depends  here on $\ep$ because of the isochrony of the 
Hamiltonian  flow gene\-ra\-ted by $p_0$.
 \item
 KAM theory (see e.g. {Ko}, \cite{AA}, \cite{Mo})  was  first introduced in 
quantum mechanics in \cite{DS} to deal with 
quasi-periodic Schr\"odinger operators. For its applications to the 
Floquet spectrum of non-autonomous  Schr\"odinger operators see 
\cite{BG} and re\-fe\-rences therein. Its first application   
to generate quantization formulas for $\hbar$ fixed  goes back to
\cite{Be} for operators in $L^2(\T^l)$  and to  \cite{Co} for 
non-autonomous perturbations of the harmonic oscillators. A uniform quantum version of the Arnold version has been obtained by Popov\cite{Po2}, within 
a quantization different from the canonical one. The related method of the  quantum 
normal forms  also yields (much less explicit) 
quantization formulas with remainders of order $O(\hbar^\infty)$, 
$O(e^{-1/\hbar^a}), 0<a<1$, $O(e^{-1/\hbar})$   
(see \cite{Sj},\cite{BGP},\cite{Po1} respectively). These formulas hold 
for a  much more 
general  class of 
symbols; however they apply only to perturbations of semi-excited levels 
(\cite{Sj, 
BGP}) or again require a quantization different from the canonical one\cite{Po1}.
\end{enumerate}
{\bf Acknowledgment} We thank Dario Bambusi for many useful comments and for pointing out an error in the first draft of this paper.
\vskip 0.5cm\noindent
\begsection{Proof of the results}
\setcounter{equation}{0}%
\setcounter{theorem}{0}%
\setcounter{proposition}{0}%
\setcounter{lemma}{0}%
\setcounter{corollary}{0}%
\setcounter{definition}{0}%
Define an analytic action $\Psi$ of $\T^l$ into
$\R^{2l}$ through the flow of $p_0$:
\begin{eqnarray}
\label{psi1}
\Psi:\T^l\times\R^{2l}\to\R^{2l},\quad 
\phi,(x,\xi) \mapsto  (x^{\prime},\xi^{\prime})=\Psi_{\phi}(x,\xi),
\end{eqnarray} 
where
\be
x^{\prime}_k:=\frac{\xi_k}{\om_k}\sin\phi_k+x_k\cos\phi_k,
\;\xi^{\prime}_k:=\xi_k\cos\phi_k-\om_kx_k\sin\phi_k. \label{psi2}
\ee
If $z:=(x,\xi)$, the  flow of initial datum $z_0$ is indeed 
$\ds z(t)=\Psi_{\om t}(z_0)$, 
 $\om t:=(\om_1 t,\ldots,\om_l t)$. 
 \par
If $f\in L^1_{loc}(\R^{2l})$, the angular
Fourier coefficient of order $k$ is defined by
$$
\tilde{f}_k (z):=\frac{1}{(2\pi)^{{l}}}\int_{\T^l}f(\Psi_{\phi}(z))e^{-i\la
k,\phi\ra}\,d\phi, \quad k\in\Z^l.
$$ 
If $f\in\C^1$ one has, as is well known
$$ 
f(\Psi_{\phi}(z))=\sum_{k\in\Z^l}\tilde{f}_k (z)e^{i\la
k,\phi\ra}\Longrightarrow f(z)=\sum_{k\in\Z^l}\tilde{f}_k (z).
$$ 
Note furthermore that $f\equiv \tilde{f}_k$ for
some fixed $k$ if and only if 
\be
\label{2.2}
 f(\Psi_{\phi}(z))=e^{i\la k,\phi\ra}f(z).
\ee 

Taking $f\in L^1(\R^{2l})$, we will consider the space Fourier transform
\be
\label{Fourier}
\widehat{f}(s):= \frac{1}{(2\pi)^{2l}}\int_{\R^{2l}}f(z)e^{-i\la s,z\ra}\,dz,
\ee 
as well the space Fourier transforms of the $\tilde{f}_k$'s:
$$
\ds \widehat{\tilde{f}_k}(s) := \frac{1}{(2\pi)^{{3l}}} \int_{\R^{2l}} 
\int_{\T^l} f(\Psi_{\phi}(z))e^{-i\la k,\phi\ra}e^{-i\la s,z\ra} d\phi\>dz.
$$
Given $\rho>0, \sigma>0$, define the norm
\be
\label{norma}
\|f\|_{\rho,\sigma}:=\sum_{k\in\Z^l}e^{\rho|k|}\int_{\R^{2l}}|
\widehat{\tilde{f}_k}(s)|
e^{\sigma |s|}\,ds.
\ee 
\begin{definition} Let $\rho>0, \sigma>0$. Then ${\cal F}_{\rho,\sigma}:=\{
f:\R^{2l}\to\C\,|\,\|f\|_{\rho,\sigma}<+\infty\}$.
\end{definition} 
\noindent
{\bf Remarks}. 
\begin{enumerate}
\item
 If $f\in {\cal F}_{\rho,\sigma}$ then $f$ is analytic on
$\R^{2l}$, and extends to a complex analytic function on a region 
${\cal B}_{\rho,\sigma}\subset\C^{2l}$ of
the form ${\cal B}_{\rho,\sigma}:=|\Im z_i|\leq a_i|\Re  z_i|$, 
with suitable $a_i$. 
\item
  $F:=Op^W_hW(f)$ is a
trace-class, self-adjoint $\hbar$-pseudodifferential operator in
$L^2(\R^l)$ if $f\in {\cal F}_{\rho,\sigma}$.  Let  $\widehat{f}(s)$ be the
Fourier transform of $f$.  Since 
$\ds \|\widehat{f}\|_{L^1}\leq
\|f\|_{\rho,\sigma}$, we have 
\be
\label{normaL^2}
\|F\|_{L^2\to L^2}\leq \int_{\R^{2l}}|\widehat{f}(s)|\,ds\equiv
\|\widehat{f}\|_{L^1}, \qquad \|F\|_{L^2\to L^2}\leq \|f\|_{\rho,\sigma}.
\ee 
\item
v We  introduce also the space ${\cal F}_{\sigma}$
of all functions
$f:\R^{2l}\to \C$ such that
$$
\|g\|_{\sigma}:=\int_{\R^{2l}}|\widehat{g}(s)|e^{\sigma |s|}\,ds < +\infty.
$$
Obviously if $f\in {\cal F}_{\sigma}$ then  $f$ is analytic on
$\R^{2l}$, and extends to a complex analytic function in the multi-strip 
 ${\cal S}:=\{z\in C^{2l}|\,|\Im  z_i|<\sigma\}$. 
\vskip 0.2cm\noindent
\item Example of $f\in {\cal F}_{\rho,\sigma}$:  $\ds 
f(x,\xi)=P(x,\xi)e^{-(|x|^2+|\xi|^2)}$, 
 $P(x,\xi)$  any polynomial.
\end{enumerate}
The starting point of the proof is represented by the first
step of the Kolmogorov iteration, and is summarized in the following
\begin{proposition}
\label{prop1}
Let $\om\in\Omega_0$. Then, for any $0<d<\rho$, $0<\delta<\sigma$:
\par\noindent 
1. 
There exists a unitary 
transformation $\ds U(\om,\ep,\hbar)=e^{i\ep W_1/\hbar}:
L^2\leftrightarrow L^2$,
$W_1=W_1^\ast$ and
$\om_1(\ep)\in[0,1]^l$ such that: 
\begin{eqnarray}
\label{passo1bis}
UH(\ep)U^{-1}=P_0(\hbar,\om_1(\ep))+\ep {\cal E}_1I+\ep^2 Q_1(\ep,\hbar)+
\ep R_1(\ep,\hbar).
\end{eqnarray}
Here:  $ {\cal E}_1=\tilde{q}_0$;  
$W_1=Op^W_h(w_1)\in \Phi_{\rho-d,\sigma-\delta}$, 
$Q_1(\ep,\hbar)=Op^W_h(q_1)\in\Phi_{\rho-d,\sigma-\delta}$ with
\be
\label{q_1}
\|w_1\|_{\rho-d,\sigma-\delta}\leq
d^{-\tau}\|q_0\|_{\rho,\sigma}\;\quad \|q_1\|_{\rho-d,\sigma-\delta}\leq
\delta^{-2}d^{-2\tau}\|q_0\|^2_{\rho,\sigma}.
\ee
2. 
$R_1(\ep)$ is a  self-adjoint semiclassical pseudodifferential
operator  of order 
$4$ such that $[R_1(\ep),P_0]=0$;  $\exists\;D_1>0$ such that, for
any eigenvector
$\psi_\alpha$ of $P_0(\om)$:
\be
\label{resto1bis}
|\la\psi_\alpha, R_1(\ep)\psi_\alpha\ra|\leq D_1(|\alpha|\hbar)^2.
\ee
3. $\forall\,K>0$ with  $\ds
(1+K^\tau)<\frac{\gamma}{\ep\|q_0\|_{\rho,\sigma}}$ $\exists$ 
$\Omega_1\subset\Omega_0$ closed and $d_1>1$ independent of $K$ such
that
\be
\label{Omega1}
|\Omega_0-\Omega_1|\leq \gamma (1+1/K^{d^1}).
\ee
Moreover if  $\om_1\in\Omega_1$ then (\ref{Diofanto}) holds with
$\gamma$ replaced by 
\be
\label{g1}
\gamma_1:=\gamma-\ep\|q_0\|_{\rho,\sigma}(1+K^\tau).
\ee
\end{proposition}
{\bf Proof}  To prove Assertion 1 we  first recall some relevant results
of
\cite{BGP}. 
\begin{lemma}[Lemma 3.6 of \cite{BGP}]
\label{omologico}
 Let $g\in\F$. Then
the  homological equation,
\be
\label{qhom}
\{p_0,w\}+{\cal N}=g,\qquad \{p_0,{\cal N}\}=0
\ee 
admits the analytic solutions 
\be
\label{soluzioni}
\label{Zg} {\cal N}:=\tilde{g}_0; \qquad w:=\sum_{k\neq
0}\frac{\tilde{g}_k}{i\la\om,k\ra},
\ee 
with the property ${\cal
N}\circ\Psi_{\phi}={\cal N}$. Equivalently, ${\cal N}$ depends only on
$I_1,\ldots,I_l$. Moreover, for any
$d<\rho$:
\be
\label{stimaom}
\|{\cal N}\|_{\rho,\sigma}\leq  \|g\|_{\rho,\sigma}; \quad 
\|w\|_{\rho-d,\sigma} \leq c_{\Psi}\frac{\|g\|_{\rho,\sigma}}{d^{\tau}}; 
\qquad
 c_{\Psi}:=\left(\frac{\tau}{e}\right)^{\tau}\frac{1}{\gamma}.
\ee 
\end{lemma} 

Given $(g,g^{\prime})\in\F$, let $\{g,g^{\prime}\}_M$ be their Moyal
bracket, defined as
$$
\{g,g^{\prime}\}_M=g\# g^{\prime}-g^{\prime}\#g,
$$
where $\#$ is the composition of  $g, g^{\prime}$ considered as Weyl
symbols. We recall that  in Fourier transform representation, used
throughout the paper, the Moyal bracket 
is (see e.g.
\cite{Fo}, $3.4$): 
\be
\label{twisted}
(\{g,g^{\prime}\}_M)^{\wedge}(s)=
\frac{2}{\hbar'}\int_{\R^{2n}}\widehat{g}(s^1)
\widehat{g^{\prime}}(s-s^1)
\sin{\left[{\hp}(s-s^1)\wedge s^1/{2}\right]}\,ds^1,
\ee
where, given two vectors $s=(v,w)$ and $s^1=(v^1,w^1)$, 
$s\wedge s^1:=\la w,v_1\ra-\la v,w_1\ra$.\par\noindent
  We also recall that $\{g,g^{\prime}\}_M=\{g,g^{\prime}\}$ if either $g$
or $g^{\prime}$ is quadratic in $(x,\xi)$. 

\begin{lemma}[Lemmas 3.1 and 3.3 of \cite{BGP}]
\label{stimeM}
 Let $g\in{\cal F}_{\sigma}$, $g^{\prime}\in{\cal 
F}_{\sigma-\delta}$. Then:
\par\noindent
1.  $\forall\,0<\delta^{\prime}<\sigma-\delta$:
\be
\label{stimaM}
\|\{g,g^{\prime}\}_M\|_{\sigma-\delta-\delta^{\prime}} \leq
\frac{1}{e^2\delta^{\prime}(\delta+\delta^{\prime})}\|g\|_{\sigma}
\|g^{\prime}\|_{\sigma-\delta}.
\ee
2.  Let $g\in\F$ and $g^{\prime}\in{\cal
F}_{\rho,\sigma-\delta}$. Then, for any positive
$\delta^{\prime}<\sigma-\delta$:
\be
\label{stimaM1}
\|\{g,g^{\prime}\}_M\|_{\rho,\sigma-\delta-\delta^{\prime}}\leq
\frac{1}{e^2\delta^{\prime}(\delta+\delta^{\prime})} \|g\|_{\rho,\sigma}\,
\|g^{\prime}\|_{\rho,\sigma-\delta}.
\ee
\end{lemma}
\vskip 0.2cm
As a simple corollary of Lemmas \ref{omologico} and \ref{stimeM}, we find:
\begin{lemma}[Lemma 3.4 of \cite{BGP}]
\par\noindent
 Let $g\in{\cal F}_{\rho,\sigma}$,  $w\in{\cal
F}_{\rho,\sigma}$.
\par\noindent
1.  Define
$$ 
g_r:=\frac{1}{r}\{w,g_{r-1}\}_{M}, \qquad r\geq 1; \;\;g_0:=g.
$$ 
Then  $g_r\in{\cal F}_{\rho,\sigma-\delta}$ for any $0<\delta<\sigma$, 
and
the following estimate holds
\be
\label{stimaindiv}
\|g_r\|_{\rho,\sigma-\delta}\leq \left(\delta^{-2}\|w\|_{\rho,\sigma}\right)^r
\|g\|_{\rho,\sigma}.
\ee
2. Let  $g\in{\cal F}_{\rho,\sigma}$, and  $w$ be the solution of the
homological equation (\ref{qhom}). Define the sequence $p_{r0}:
r=0,1,\ldots$ as follows:
$$ 
p_{00}:=p_0; \qquad p_{r0}:=\frac{1}{r}\{w,p_{r-10}\}_M, \;r\geq 1.
$$ 
Then, for any $0<d<\rho, 0<\delta <\sigma$,   $p_{r0}\in{\cal
F}_{\rho-d,\sigma-\delta}$ and fulfills the following estimate
$$
\|p_{r0}\|_{\rho-d,\sigma-\delta} \leq
2\left(\delta^{-2}\|w\|_{\rho-d,\sigma}\right)^{r-1}\|g\|_{\rho-d,\sigma},
\;\;k\geq 1.
$$
\end{lemma} 
\vskip 0.2cm\noindent
{\bf Proof of Proposition \ref{prop1}}
\par\noindent
With $\ds U_1=e^{i\ep W_1/\hbar}$, $W_1$ continuous and self-adjoint, we
have in general: 
\begin{eqnarray}
\label
{U1}
U_1(P_0+\ep Q_0)U_1^{-1}=P_0+\ep P_1+\ep ^2 Q_1, \qquad\qquad 
\\
\label{P1}
P_1:= Q_0+[W_1,P_0]/i\hbar, \qquad\qquad \qquad\qquad 
\\
\label{Q1}
Q_1:=\ep^{-2}\left(U_1(P_0+\ep
Q_0)U_1^{-1}-P_0-\ep (Q_0+[W_1,P_0]/i\hbar)\right).
\end{eqnarray}
We start by looking for $W_1\in\F$ such that the first order term yields an
operator $N_1\in\F$ commuting with $P_0$:
\be
\label{qhom1}
Q_0+[W_1,P_0]/i\hbar =N_1,\quad [N_1,P_0]=0.
\ee
Denoting by $w_1$, ${\cal N}_1$ the (Weyl) semiclassical symbols of $W_1$,
$N_1$, respectively, eq.(\ref{qhom1}) is equivalent to a classical
homological equation in $\F$
\be
\label{chom1}
\{p_0,w_1\}_M+{\cal N}_1=q_0, \qquad \{p_0,{\cal N}_1\}_M=0.
\ee
However $p_0$ is quadratic in $(x,\xi)$. Therefore the Moyal bracket
$\{p_0,w_1\}_M$ coincides with the Poisson bracket $\{p_0,w_1\}$ and the
above equation becomes
\be
\label{chom2}
\{p_0,w_1\}+{\cal N}_1=q_0, \qquad \{p_0,{\cal N}_1\}=0.
\ee
The existence of
$w_1\in{\cal F}_{\rho-d,\sigma}
$,
${\cal N}_1\in\F$ with the stated properties now follows by direct
application of Lemma
\ref{omologico}.

We now prove the second estimate in (\ref{q_1}). We have:
$$
Q_1=\int_0^1\!\!\int_0^se^{is_1\ep W_1/\hbar}[[P_0+\ep
Q_0,W_1],W_1]e^{-is_1\ep W_1/\hbar}\,ds_1ds,
$$
and we can estimate
$$
\|[[P_0+\ep Q_0,W_1],W_1]\|_{L^2\to L^2}\leq \|\{\{p_0+\ep
q_0,w_1\}_M,w_1\}_M\|_{\rho-d,\sigma-\delta}.
$$
It follows, by Lemma 2.3 and Lemma 2.1, that
\begin{eqnarray*}
\|Q_1\|_{L^2\to L^2} \leq  \|\{\{p_0+\ep
q_0,w_1\}_M,w_1\}_M\|_{\rho-d,\sigma-\delta}\leq
\delta^{-2}d^{-2\tau}\|q_0\|_{\rho,\sigma}^2.
\end{eqnarray*}
This proves the second estimate of (\ref{q_1}). 

To prove the Assertion 2 set:
\begin{eqnarray}
\label{N1}
{\cal E}_1&:=&{\cal N}_1(0); \quad
\om_1(\ep)=\om+\ep(\nabla_I{\cal N}_1)(0),
\\
\label{calR1}
 {\cal R}_1(I,\ep)&=&
{\cal N}_1(I)-\la(\nabla_I{\cal N}_1)(0),I\ra-{\cal E}_1,
\end{eqnarray}
and define
\be
\label{R1}
 {R_1}(\ep):=Op^W_h({\cal R}_1(I,\ep)).
\ee
Then clearly ${R_1}(\ep)$ is a self-adjoint semiclassical, tempered 
pseudodifferential operator of order $4$, vanishing to 4-th order
at the origin, and with the property
$[R_1(\ep),P_0]=0$. Therefore formula (\ref{resto1bis}) follows directly
by Proposition A.1. 

As far as Assertion 3 is concerned, set:
\begin{eqnarray}
\label{Tk}
{\cal T}_k(\alpha)&:=&\{\om\in [0,1]^k: |\la\om,k\ra|\leq \alpha\},
\\
\label{Pi1}
\Om_1&:=&\Om_0 -\bigcup_{|k|\geq
K}{\cal T}_k\left(\frac{\gamma_1}{|k|^\tau}\right).
\end{eqnarray}
As in \cite{BG}, Lemma 5.6, we have:
$$
|{\cal T}_l(\alpha)|\leq \frac{4}{k}\alpha.
$$
Hence if $\tau >l-1$ we can write
$$
\left|\bigcup_{|k|\geq
K}{\cal T}_k\left(\frac{\gamma_1}{|k|^\tau}\right)\right|\leq
\sum_{|k|\geq K}\frac{\gamma_1}{|k|^{\tau+1}} <\frac{\gamma_1}{K^{d_1}}.
$$
Since $\ds |\la\om_1(\ep),k\ra|\geq \gamma_1/|k|^\tau$ by construction
when $|k|\leq K$, the proposition is proved.
\vskip 1.0cm\noindent
\section{Iteration}
\setcounter{equation}{0}%
\setcounter{theorem}{0}%
\setcounter{proposition}{0}%
\setcounter{lemma}{0}%
\setcounter{corollary}{0}%
\setcounter{definition}{0}%
 The above result represents the
starting point for the iteration. To ensure convergence, we first preassign the
values of the parameters involved in the iterative estimates. Keeping
$\ep$, $K$, $\gamma$, $\rho$ and $\sigma$ fixed define, for $p\geq 1$:
\begin{eqnarray}
\label{iter1}
 \sigma_p &:=&\frac{\sigma}{4p^2},\quad
s_p:=s_{p-1}-\sigma_p,\quad  
\rho_p:=\frac{\rho}{4p^2},\quad
r_p:=r_{p-1}-\rho_p,
\\
\label{iter2} 
 \gamma_p&:=&\gamma_{p-1}-\frac{4\ep_p}{1+K_p^{\tau}}, \quad 
K_p:=pK.
\end{eqnarray}
where $\ep_p$ is defined in (\ref{epsp}) below. The initial values of the parameter sequences are chosen as follows:
\be
\label{iter3}
\gamma_0:=\gamma;\quad s_0:=\sigma;\quad r_0:=\rho, \quad \ep_0=0.
\ee
 We then have:
\begin{proposition}
\label{prop2}
let $\om\in\Omega_0$. There exist $\ep^\ast(\gamma)>0$ and,
$\forall\,p\geq 1$, a closed set $\Omega_p^\gamma\subset \Omega_0$  such
that, if $|\ep|<\ep^\ast(\gamma)>0$ and $\om_p(\hbar;\ep)\in\Omega_p^\gamma$:
\par\noindent
1. One can construct  two sequences of unitary  transformations
$\{X_p\}$,
$\{Y_p\}$ in
$L^2(\R^l)$ with the property 
\begin{eqnarray}
\label{equivk}
X_p(P_0(\om)+\ep
Q_0)X_p^{-1}=\qquad\qquad\qquad\qquad\qquad\qquad
\\
\nonumber
P_0(\om_p(\hbar;\ep))+\ep{\cal E}_p(\hbar;\ep)I+e^{2^p}Q_p+
\\
\nonumber
\ep^{2^p}R_p(\hbar;\ep)+
\ep\sum_{s=2}^{p}Y_sR_{s-1}(\hbar)Y_s^{-1}\ep^{2^{s-2}}.
\end{eqnarray}
\par\noindent 
2. $X_p$ and $Y_p$ have the form
\begin{eqnarray}
\label{unitarie}
X_p=U_1U_2\cdots U_p;\\
Y_s=U_pU_{p-1}\cdots U_s.
\end{eqnarray}
Here $\ds U_p(\om,\ep,\hbar)=\exp{[i\ep^{2^{p-1}} W_p/\hbar}]:
L^2\leftrightarrow L^2$,
$W_p=W_p^\ast$
\begin{eqnarray}
\label{wp}
W_p=Op^W_h(w_p)\in \Phi_{r_p,s_p}, \quad 
Q_p(\ep,\hbar)=Op^W(q_p)\in\Phi_{r_p,s_p},
\\
\label{stimewp}
\|w_p\|_{r_p,s_p}\leq
\rho_p^{-2\tau}\|q_{p-1}\|_{r_{p-1},s_{p-1}}\;\quad
\|q_p\|_{r_p,s_p}\leq
\rho_p^{-2\tau}\sigma_p^{-2}\|q_{p-1}\|^2_{r_{p-1},s_{p-1}},
\\
{\cal E}_p(\hbar;\ep)=\sum_{s=0}^{p}{\cal 
N}_s(\hbar)\ep^{2^{s}}, \quad {\cal
N}_s(\hbar)=(\tilde{q}_s)_0(\hbar).\qquad
\end{eqnarray}
\par\noindent
3. 
$R_s(\ep)$ is a self-adjoint semiclassical pseudodifferential operator 
of order $4$; $[R_s(\ep),P_0]=0$; 
there exist $D_{p}>0, \overline{D}_{p}>0$ such that, for any eigenvector
$\psi_\alpha$ of $P_0(\om)$:
\begin{eqnarray}
\label{resto1ter}
|\la\psi_\alpha, R_p(\ep)\psi_\alpha\ra|\leq \ 
D_{p}(|\alpha|\hbar)^2,\qquad\quad
\\
\label{resto1ter2}
|\la\psi_\alpha,
\sum_{s=2}^{p}Y_sR_{s-1}Y_s^{-1}\ep^{2^{s-2}}\psi_\alpha\ra|\leq 
\overline{D}_{p}(|\alpha|\hbar)^2.
\end{eqnarray}
4. $\forall\,K_{p-1}>0$ such that 
\be
\label{Kp}
(1+K_{p-1}^\tau)<\frac{\gamma_{p-1}}{\ep\|q_{p-1}\|_{r_{p-1},s_{p-1}}},
\ee
$\exists$ 
$\Omega_p\subset\Omega_{p-1}$ closed and $d_p>1$ independent of $K_p$ such
that
\be
\label{Omegap}
|\Omega_p-\Omega_{p-1}|\leq \frac{\gamma_{p-1}} {1+1/(K_{p-1})^{d_p}}.
\ee
Moreover if  $\om_p(\ep)\in\Omega_p$ then (\ref{Diofanto}) holds with
$\gamma$ replaced by 
\begin{eqnarray}
\label{gp}
\gamma_p&:=&\gamma_{p-1}-\ep_p(1+K_{p-1}^\tau)
\\
\label{epsp}
\ep_p&:=&\ep^{2^{p-1}}\|q_{p-1}\|_{r_{p-1},s_{p-1}}
\end{eqnarray}
\end{proposition}
{\bf Proof}
\par\noindent
We proceed by induction. For $p=1$ the assertion is true because we can
take
$W_1$,
$Q_1$,
$R_1$,
$\om_1$,
$\Om_1^\ep$,
$K_1$ as in   Proposition \ref{prop1}. To go from step $p-1$ to step $p$
we consider the operator 
\begin{eqnarray*}
X_{p-1}(P_0(\om)+\ep Q_0)X_{p-1}^{-1}:=\qquad\qquad\qquad\qquad
\\
P_0(\om_{p-1}(\hbar;\ep))+\ep{\cal
E}_{p-1}(\hbar;\ep)I+e^{2^{p-1}}Q_{p-1}
\\
+\ep^{2^{p-1}}R_{p-1}(\hbar;\ep)+
\ep\sum_{s=2}^{{p-1}}Y_sR_{s-1}(\hbar)Y_s^{-1}\ep^{2^{s-2}}.
\end{eqnarray*}
We have to determine and estimate the unitary map $U_p$ transforming
it into the form (\ref{equivk}) via the definitions (\ref{unitarie}).
With $\ds U_p=e^{i\ep W_p/\hbar}$, $W_p$ continuous and self-adjoint, we
have at the $p$-th iteration step 
\begin{eqnarray*}
U_p(P_0(\om_{p-1}+\ep^{2^{p-1}} Q_{p-1})U_p^{-1}=P_0(\om_p)+\ep^{2^{p-1}}P_p+\ep^{2^p} Q_p,
\qquad\qquad 
\\
P_p:= Q_{p-1}+[W_p,P_0]/i\hbar, \qquad\qquad \qquad\qquad 
\\
Q_p:=\ep^{-2}\left( U_p(P_0(\om_{p-1})+\ep
Q_0)U_1^{-1}-P_0(\om_{p-1})-\ep (Q_{p-1}+[W_p,P_0]/i\hbar)\right).
\end{eqnarray*}
(the explicit dependence of the frequencies on $(\hbar,\ep)$ has been omitted). We will look therefore  for
$W_p\in\Phi_{r_p,s_p}$ and an operator
$N_p\in
\Phi_{r_p,s_p}$  such that
\be
\label{qhomp}
Q_p+[W_p,P_0]/i\hbar =N_p,\quad [N_p,P_0]=0.
\ee
Denoting $w_p$, ${\cal N}_p$ the (Weyl) semiclassical symbols of $W_p$,
$N_p$, respectively, eq.(\ref{qhomp}) is again equivalent to the classical
homological equation in $\F$
$$
\{p_0,w_p\}_M+{\cal N}_p=q_p, \qquad \{p_0,{\cal N}_p\}_M=0
$$
which once more becomes
$$
\{p_0,w_p\}+{\cal N}_p=q_p, \qquad \{p_0,{\cal N}_p\}=0.
$$ 
The existence of
$w_p\in{\cal F}_{r_p,s_p}$,
${\cal N}_p\in{\cal F}_{r_p,s_p}$ with the stated properties now follows
by direct application of Lemma
\ref{omologico}. Expanding ${\cal N}_p$ as in the proof of
Proposition \ref{prop1} and taking into account the definitions
(\ref{unitarie}) we immediately check that $\ds X_pX_{p-1}(P_0(\om)+\ep
Q_0)X_{p-1}^{-1}X_p$ has the form (\ref{equivk}). The estimate
of
$Q_p$ and the small denominator estimates follow by exactly the same
argument of Proposition 2.1. The estimate (\ref{resto1ter}) is proved
exactly as (\ref{resto1bis}). It remains to prove the estimate
(\ref{resto1ter2}).  By the inductive assumption, it is enough to prove the
existence of $D^\prime_p>0$ such that
$$
|\la \psi_\alpha,U_pR_{p-1}U_{p}^{-1}\psi_\alpha\ra|\leq 
D^\prime_p(|\alpha|\hbar)^2.
$$
We only have to prove that the operator
$U_pR_{p-1}U_{p}^{-1}$ is an $\hbar$-pseudo\-differential operator of order
$4$ fulfilling the   hypotheses of Proposition A.1, assuming  by the
inductive argument the validity of these properties for  
$R_{p-1}$. On the other hand, $\ds U_p=\exp{(i\ep^{2^{p-1}}W_p/\hbar)}$, and $W_p$ is an 
$\hbar$-pseudo\-differential operator of order
$0$.  We can therefore apply the semiclassical Egorov theorem 
(see e.g. \cite{Ro}, 
Chapter 4) to assert that $U_pR_{p-1}U_{p}^{-1}$ is again an 
$\hbar$-pseudo\-differential operator. Denote 
$\sigma (x,\xi;\ep;\hbar)$ the Weyl symbol of $U_pR_{p-1}U_{p}^{-1}$, and 
consider its expansion
$$
\sigma(x,\xi;\ep;\hbar)=\sigma_0(x,\xi;\ep)+\sum_{j=2}^M\hbar^j\sigma_j(x,\xi;\ep
)+O(h^{M+1}).
$$
It is clearly enough to prove that the principal symbol $\sigma_0(x,\xi;\ep)$ has 
order $4$. Denote by
$$
\phi(x,\xi;\ep):=\exp{[\ep^{2^p}{\cal L}_{w_p}]}(x,\xi)
$$ 
the Hamiltonian flow  on $\R^{2l}$ generated by the 
Hamiltonian vector field $\ds {\cal L}_{w_p}$ at time $\ds \ep^{2^p}$;
here $w_p^0(x,\xi)$ is the  principal symbol of $W_p$. Then
$\sigma_0(x,\xi;\ep)={\cal  R}_{p-1}^0(\phi(x,\xi;\ep))$ where ${\cal
R}_{p-1}^0(x,\xi)$ is in turn the  principal symbol of $R_{p-1}$. Now 
$$
\phi(x,\xi;\ep)=(x+\int_0^{\ep^{2^p}}\nabla_\xi w_p(x,\xi;\eta)\,d\eta,
\xi-\int_0^{\ep^{2^p}}\nabla_x w_p(x,\xi;\eta)\,d\eta).
$$
By Assumption A2 and the inductive hypothesis we know that
$w_p(z)=O(|z|^2)$ as $|z|\to 0$. Hence we can write $\phi(z)=z+\epsilon
r(z)$ where $r(z)=O(z), z\to 0$. 
 This concludes the proof of Proposition 3.1. 
\vskip 0.3cm
\noindent
{\bf Proof of Theorem \ref{mainth}}
\vskip 0.2cm\noindent
Applying the estimates on $q_p$ in Propositions \ref{prop1} and \ref{prop2}  
iteratively, we have
\begin{equation}
\label{qp}
\|q_p\|_{r_p,s_p}\leq \left(\frac{4p^2}{\rho}\right)^{2\tau p}\cdot 
\left(\frac{4p^2}{\sigma}\right)^{2p}\|q_0\|^{2^p},
\end{equation}
whence
\begin{equation}
|\ep|^{2^p}\|Q_p\|_{L^2\to L^2}\leq 
|\ep|^{2^p}(4p^2)^{2p(\tau+1)}\rho^{-2\tau p} \sigma^{-2p} \|q_0\|^{2^p}
\to 0\quad \hbox{as }p\to\infty,
\end{equation}
for all $|\ep|\leq \ep^\ast$ provided $\ep^\ast>0$ is small enough.  
At the $p$-th iteration the frequency is given by
\begin{equation}
\label{omk}
\om_p(\hbar;\ep)=\om +\sum_{s=1}^p\nabla_I{\cal 
N}_s(\hbar)\ep^{2^s}.
\end{equation}
Since  $\ds \|\nabla_z f(z)\|_{\rho-d,\sigma-\delta}\leq 
\frac{1}{d\delta}\|f(z)\|_{\rho,\sigma}$, by (\ref{qp}) we have
\begin{equation}
\label{Nk}
\sum_{s=1}^p|\nabla_I{\cal N}_s(\hbar)\ep^{2^s}|\leq
\sum_{s=1}^p |\ep|^{2^s}(4s^2)^{2s(\tau+1)}\rho^{-2\tau s} 
\sigma^{-2s} \|q_0\|^{2^s}.
\end{equation}
Hence the series (\ref{omk}) converges as $p\to \infty$
for $|\ep|<\ep^\ast$ if $\ep^\ast $ is small enough, uniformly with respect 
to $\hbar)\in [0,h^\ast]$. In the same way, 
the estimate  (\ref{qp}) entails,  
by the definition (\ref{gp}), 
the existence of $\ds \lim_{p\to\infty}\gamma_p:=\gamma_\infty$. 
Let $\om(\hbar;\ep) := 
\lim_{p\to\infty} 
\om_p(\hbar\;\ep)$.  Then $\om(\hbar;\ep)$ is diophantine 
with constant $\gamma_\infty$ by Proposition \ref{prop2}.  
In the same way:
$$
{\cal E}(\hbar;\ep)=\sum_{s=1}^\infty{\cal 
N}_s(\hbar)\ep^{2^s}, \quad 
|\ep|<\ep^\ast.
$$
Finally, let ${\cal R}(\alpha\hbar,\ep)$ be an asymptotic sum of the power series  $\ds \sum_{s=2}^{\infty}Y_sR_{s-1}Y_s^{-1}\ep^{2^{s-2}}$. Then the validity 
of (\ref{resto}) follows by its validity term by term. This concludes
the proof of  Theorem \ref{mainth}.
\par\noindent
{\bf Proof of Corollary 1.1}
\par\noindent
It is enough to illustrate the specialization of the argument of 
Propositions 2.1 and 3.1 to the $\hbar=0$ case. 
Denoting by  $\ds e^{\ep {\cal L}_{w_1}}$ the canonical 
flow at time $\ep$ generated by the  Hamiltonian vector field 
generated by the symbol $w_1$, we have:
\begin{eqnarray}
\label{U1c}
e^{\ep {\cal L}_{w_1}}(p_0+\ep q_0)(x,\xi)=(p_0+\ep p_1+\ep ^2 q_1^0)(x,\xi), 
\qquad\qquad 
\\
\label{P1c}
p_1:= q_0+\{w_1,p_0\}, \qquad\qquad \qquad\qquad 
\\
\label{Q1c}
q_1^0:=\ep^{-2}\left( e^{\ep {\cal L}_{w_1}}(p_0+\ep q_0)(x,\xi)-
p_0-\ep (q_0+\{w_1,p_0\})\right).
\end{eqnarray}
Remark that $\ds e^{\ep{\cal L}_{w_1}}(p_0+\ep q_0)(x,\xi)$ is the 
principal symbol of $U_1(P_0+\ep Q_0)U_1^{-1}$ by the 
semiclassical Egorov theorem;  $p_1$ is  the full, and hence principal,  
symbol of
 $P_1$ because $p_0$ is quadratic. Likewise, $q_1^0$ is  
the principal symbol  of $Q_1$. Hence the classical 
definitions (\ref{U1c},\ref{P1c},\ref{Q1c}) correspond to 
the principal symbols of the semiclassical pseudodifferential 
operators $U_1(P_0+\ep Q_0)U_1^{-1}$, $P_1$, $Q_1$ defined in 
(\ref{U1},\ref{P1},\ref{Q1}). Therefore we can take over the 
homological equation (\ref{chom2}) and apply Lemma \ref{omologico}
 once more. This yields the same $w_1$ and ${\cal N}_1$ of 
Proposition 2.1. To prove the estimate  (\ref{q_1}) for $q_1^0$ we write 
$$
q_1^0=\int_0^1e^{s\ep {\cal L}_{w_1}}\{\{p_0+\ep
q_0,w_1\},w_1\}\,ds
$$
Now as in \cite{BGGS}, Lemma 1, note that if $|\ep |<\ep^\ast$ and 
$z=(x,\xi)\in {\cal B}_{\rho-d,\sigma-\delta}$ then $e^{s{\cal L}_{w_1}}z\in
{\cal B}_{\rho,\sigma}$ for $0\leq s\leq 1$ because (Lemma 2.1) 
$\ds \ep\|\nabla w_1\|_{\rho-d,\sigma} \leq \ep (\tau/e)c_\psi 
d^{-\tau}\|q_0\|_{\rho,\sigma}$. 
Therefore we can apply Lemma 2.3, valid a fortiori 
for the Poisson bracket, and, as in the 
proof of Proposition 2.1, get the 
estimate corresponding to the second one of (\ref{q_1}): 
\begin{eqnarray}
\label{stimaq10}
\|q_1^0\|_{\rho-d,\sigma-\delta} \leq  \|\{\{p_0+\ep
q_0,w_1\},w_1\}\|_{\rho-d,\sigma-\delta}\leq \delta^{-2}
d^{-2\tau}\|q_0\|_{\rho,\sigma}^2.
\end{eqnarray}
Now, writing: 
\begin{eqnarray}
\psi^1_{\ep}(x,\xi)&=&e^{\ep {\cal L}_{w_1}}(x,\xi), \quad 
\label{N1c}
{\cal E}_1:={\cal N}_1(0); 
\\
\label{omega1}
\om_1(\ep)&=&\om+\ep(\nabla_I{\cal N}_1)(0),
\\
\label{R1c}
\tilde{\cal R}_1(I,\ep)&=&
{\cal N}_1(0)-\la(\nabla_I{\cal N}_1)(0),I\ra-{\cal E}_1,
\end{eqnarray}
we can sum up the above argument by writing (compare with (\ref{passo1bis}))
\be
\label{passo1c}
\psi^1_{\ep}\circ (p_0+\ep q_0)={\cal 
E}_1+\la\om_1(0;\ep),I\ra+\ep^2q_1(I,\phi)+\ep {\cal R}_1^0(I,\ep)
\ee
where ${\cal R}^0_1$ is the principal symbol of $R_1$. 
 Morover, Assertion 3 of Proposition 2.1 holds  without change. 
\newline
Let us now specialize the iterative  argument of Proposition 3.1. First, the 
parameters defined in (\ref{iter1},\ref{iter2},\ref{iter3}) remain unchanged. 
Then:
\newline 
1. The construction of the  two sequences of canonical  transformations
\begin{eqnarray}
\chi^p_{\ep}&=&\psi^1_{\ep}\circ\psi^2_{\ep}\cdots\circ  \psi^p_{\ep}, 
\quad p=1,2,\ldots
\\
\zeta^s_{\ep}&=&\psi^p_{\ep}\circ\psi^{p-1}_{\ep}\cdots\circ  \psi^s_{\ep}, 
\quad p=1,2,\ldots
\\
\psi^s_{\ep}(x,\xi)&=&e^{\ep {\cal L}_{w^0_s}}(x,\xi)
\end{eqnarray}
such that 
\begin{eqnarray}
\label{equivkc}
 \psi^p_{\ep,I_0}\circ (p_0+\ep q_0)=\qquad\qquad\qquad\qquad\qquad
\\
\nonumber
\la\om_p(0,\ep),I\ra+{\cal E}_p(\ep)+e^{2^p}q^0_p+\ep^{2^p}{\cal R}^0_p+
\ep\sum_{s=2}^{p}\psi^s_{\ep}\circ {\cal R}_{s-1}^0\ep^{2^{s-2}}.
\end{eqnarray}
follows as in the above argument valid for $p=1$. 
Here $w_s^0$, $q^0_p$, ${\cal R}^0_s$ are the principal symbols 
of the semiclassical pseudodifferential operators $W_s$, $Q_p$ and $R_s$, 
once reexpressed on the $(x,\xi)$ canonical variables via, 
with $\om_p$ in place of $\om_1$. Morover:
\begin{eqnarray}
{\cal E}_p(\ep)=\sum_{s=0}^{p}{\cal 
N}_s(0)\ep^{2^{s}}, \quad {\cal
N}_s(0)=(\tilde{q}_s^0)_0(0).\qquad
\\
\om_p(\ep)=\om+\sum_{s=0}^{p}\om_s(0)\ep^{2^{s}},\quad 
\om_s(0)=\nabla_I{\cal N}_s(0)
\end{eqnarray}
2.  The estimates (\ref{stimewp}) are a fortiori valid with $w^0_p$, 
$q^0_p$ in place of $w_p$, $w_p$; as a consequence, 
(\ref{Omegap}) holds unchanged together with the definitions 
(\ref{Kp},\ref{gp},\ref{epsp}). 
Hence the uniform estimate (\ref{qp}) allows  us to 
set $\hbar=0$ in (\ref{omk},\ref{Nk}).
\par\noindent 
3. Finally, 
remark that ${\cal R}^0_s(I)=O(I^2), s=1,\ldots,p$. Now 
 the estimate $\psi_s^\ep{\cal R}_s(I)=O(I^2)$ as $I\to 0$ follows by exactly 
the same argument of Proposition 3.1  after 
rexpression on the canonical variables $(x,\xi)$.
\vfill\eject
\noindent
{\bf\Large Appendix}
\vskip 0.3cm\noindent
\setcounter{equation}{0}%
\setcounter{theorem}{0}%
\setcounter{proposition}{0}%
\setcounter{lemma}{0}%
\setcounter{corollary}{0}%
\setcounter{definition}{0}%
To establish the remainder estimate (\ref{resto}) the key fact is that vanishing
of a symbol  at the origin $(x,\xi)=0$ implies bounds on harmonic
oscillator matrix elements that are uniform in $\hbar$.  No analyticity
of the symbol is required for this result, so we will state and prove it
in somewhat greater generality, using the following  semiclassical symbol
class defined in Shubin \cite{Sh}:
$$
\Sigma^{m,\mu} = \{f\in C^\infty({\R}^{2l}\times (0,\epsilon]):\; 
|\partial^\gamma_z
f(z,\hbar)| \le C_\gamma \langle z\rangle^{m-|\gamma|} \hbar^{\mu}\},
$$
where $z = (x,\xi)$, here considered a real variable, and $\langle z\rangle = 
\sqrt{1+|z|^2}$.   
For future reference we note that
Proposition A.2.3 of \cite{Sh}  gives the result:
$$
\forall f \in \Sigma^{0,\mu}, \quad \Vert Op^{W}_\hbar(f) \Vert_{L^2}
\le C(f) \hbar^\mu,
\eqno{(A.1)}
$$
for all $\hbar\in(0,\epsilon]$.

The matrix elements in question are most easily computed in Bargmann 
space,
with the remainder operator written as a Toeplitz operator.  Since these are 
anti-Wick ordered, we first must consider the translation from Weyl symbols to
anti-Wick (for these notions, see e.g. \cite{BS}). 
Denoting by $Op^{AW}_\hbar(f)$ the anti-Wick quantization of a symbol 
$f\in \Sigma^{m,\mu}$, the correspondence is given by the action of the heat
kernel on the symbol:
$$
Op^{AW}_\hbar(f) = Op^{W}_\hbar(e^{\hbar\Delta/4} f),
\eqno{(A.2)}
$$
where $\Delta = \Delta_z = \partial_x\cdot\partial_x 
+ \partial_\xi\cdot\partial_\xi$.
To begin, we show that the Weyl symbol of an anti-Wick operator is given by
formal expansion of the heat kernel up to a remainder.
\vskip 0.2cm\noindent
{\bf Lemma A1}
{\it 
For $f,g\in \Sigma^{m,\mu}$,  suppose that $Op^{AW}_\hbar(g) = Op^{W}_\hbar(f)$.
Then for all $n\ge 1 $,}
$$
f - \sum_{k=0}^{n-1}\frac{1}{k!} \left(\frac{\hbar}{4}\Delta\right)^k g \in 
\Sigma^{m-2n,\mu+n}.
$$
{\bf Proof.}
According to (A.2), 
$$
f(z,\hbar) = \frac{1}{(\pi\hbar)^l} \int e^{-|z-w|^2/\hbar} g(w) dw.
$$
In this expression we will expand $g(w)$ in a Taylor series centered at $w=z$:
$$
g(w,\hbar) = \sum_{|\alpha|<2n} \frac{1}{\alpha!} \partial^\alpha g(z,\hbar)
(w-z)^\alpha + r(w,z,\hbar),
$$
where
$$
r(w,z) = \sum_{|\alpha|=2n} c'_\alpha (w-z)^\alpha \int_0^1 (1-t)^{2n-1}
\partial^\alpha g(z+t(w-z)) \>dt.
$$
Thus,
$$
f(z,\hbar) = \sum_{|\alpha|<2n} c_\alpha \partial^\alpha g(z,\hbar) + r(z,\hbar),
$$
where 
$$
c_\alpha = \frac{1}{(\pi\hbar)^l} \frac{1}{\alpha!} \int w^\alpha 
e^{-|w|^2/\hbar}\>dw,
$$
and 
$$
r(z,\hbar) = \sum_{|\alpha|=2n} c''_\alpha \hbar^{-l} \int\int_0^1 (w-z)^\alpha 
e^{-|z-w|^2/\hbar} (1-t)^{2n-1} \partial^\alpha g(z+t(w-z)) \>dt\>dw.
$$
Note that $c_\alpha = 0$ for $|\alpha|$ odd, and for any integer $k$
$$
\sum_{|\alpha| = 2k} c_\alpha \partial^\alpha g = \frac{1}{k!} 
\left(\frac{\hbar}{4}\Delta\right)^k g.
$$
The lemma is thus reduced to the claim that $r(z,\hbar) \in \Sigma^{m-2n,\mu+n}$.

To see this, we change variables by $w' = (w-z)/\sqrt{\hbar}$ to write
$$
r(z,\hbar) = \sum_{|\alpha|=2n} c''_\alpha \hbar^{n} \int\int_0^1 w^\alpha 
e^{-|w|^2} (1-t)^{2n-1} \partial^\alpha g(z+tw\sqrt{\hbar}) \>dt\>dw.
$$
We must estimate the derivatives:
$$
\partial^\gamma r(z,\hbar) = \sum_{|\alpha|=2n} c''_\alpha \hbar^{n} 
\int\int_0^1 w^\alpha 
e^{-|w|^2} (1-t)^{2n-1} \partial^\beta g(z+tw\sqrt{\hbar}) \>dt\>dw,
$$
where $|\beta| = 2n+|\gamma|$.  This integral for $\partial^\gamma r$ 
we then split into two pieces according to the domain of the $w$-integral,
$I'_{\alpha,\beta}:|w|<|z|/2$
and $I''_{\alpha,\beta}:|w|>|z|/2$.  The assumption 
$g \in \Sigma^{m,\mu}$ implies an estimate
$$
|I'_{\alpha,\beta}| \le C \langle z\rangle^{m-2n-|\gamma|} \hbar^{n+\mu}.
\eqno{(A.3)}
$$
The second term is taken care of by the exponential factor in $|w|$:
$$
|I''_{\alpha,\beta}| < C_l \hbar^l \langle z\rangle^{-l}, \quad\forall l.
$$
Therefore $\partial^\gamma r$ satisfies an estimate of the form 
(A.3)
for any $\gamma$, and hence $r \in \Sigma^{m-2n,\mu+n}$.\quadratino
\vskip 0.2cm\noindent

Our application of Lemma A.1 will be specifically 
to operators of order 4:
\vskip 0.2cm\noindent
{\bf Lemma A.2}
 {\it For $g \in \Sigma^{4,0}$,
$$
Op^{W}_\hbar(g) = Op^{AW}_\hbar(g) - \frac{\hbar}{4}  Op^{AW}_\hbar(\Delta g)
+ R(\hbar),
$$
where $\Vert R(\hbar)\Vert_{L^2} \le C\hbar^2$ .}
\vskip 0.2cm\noindent
{\bf Proof.}
Let $\sigma(A)$ denote the Weyl symbol of the $\hbar$-pseudodifferential 
operator $A$.  Applying Lemma A.1 with $n=2$ gives
$$
\sigma(Op^{AW}_\hbar(g)) = g + \frac{\hbar}{4} \Delta g + r_1,
$$
and
$$
\frac{\hbar}{4} \sigma(Op^{AW}_\hbar(\Delta g)) = \frac{\hbar}{4} \Delta g + r_2,
$$
where $r_1, r_2\in \Sigma^{0,2}$. Noting that
$$
Op^{W}_\hbar(g) - Op^{AW}_\hbar(g) + \frac{\hbar}{4} \sigma(Op^{AW}_\hbar(\Delta 
g))
 = Op^{W}_\hbar(r_1-r_2),
$$
the bound on $R(\hbar)$ follows from (A.1).\quadratino
\vskip 0.2cm\noindent

The point of introducing anti-Wick symbols is to exploit 
the Bargmann space representation of the harmonic oscillator.  The Bargmann space
is (see e.g. \cite{BS})
$$
\mathcal{H}_\hbar = L^2_{hol}({\C}^l, e^{-|z|^2/\hbar} \>dzd\bar z).
$$
  The Bargmann transform is an
isomorphism
$\mathcal{B}: L^2({\R}^l) \to \mathcal{H}_\hbar$, defined so as to
intertwine anti-Wick operators with Toeplitz operators:
$$
\mathcal{B}\circ Op^{AW}_\hbar(f)\circ\mathcal{B}^{-1} = T_\hbar(f).
$$
The Toeplitz operator $T_\hbar(f):\mathcal{H}_\hbar\to \mathcal{H}_\hbar$ 
is defined for $f\in\Sigma^{m,\mu}$ by 
$$
T_\hbar(f) = \Pi_\hbar M(f),
$$ 
where $M(f)$ denotes the multiplication operator on 
$L^2({\C}^l, e^{-|z|^2/\hbar} \,dzd\bar z )$
(identifying ${\R}^{2l} = {\C}^l$ by $z = x+i\xi$), 
and $\Pi_\hbar:L^2({\C}^l, e^{-|z|^2/\hbar} \>dzd\bar z) \to \mathcal{H}_\hbar$
is orthogonal projection onto the holomorphic subspace.

The main result of this Appendix is the following matrix element estimate:
\vskip 0.2cm\noindent
{\bf Proposition A.1}
{\it Let $\{\psi_\alpha\}$ be the normalized eigenstates of the standard
harmonic  oscillator on $L^2({\R}^{l})$.  Suppose $f\in \Sigma^{4,0}$
satisfies
$$
f(z,\hbar) = \sum_{|\gamma|=4} z^\gamma g_\gamma(z,\hbar),
$$
where $\sup|\partial^\beta g_\gamma|\le M$ for all $|\beta|\le 2$.
Then 
$$
|\langle \psi_\alpha, Op^W_\hbar(f) \psi_\alpha\rangle| \le
C M (|\alpha| \hbar)^{2} 
$$
for all $\alpha, \hbar$, where $C$ depends only on the dimension.}
\vskip 0.2cm\noindent
{\bf Proof.}
 Under the Bargmann transform the harmonic oscillator eigenstates have a particularly
convenient form:
$$
(\mathcal{B}^{-1} \psi_\alpha) (z) = (\pi^l \hbar^{|\alpha|+l} \alpha!)^{-1/2}
\cdot z^\alpha.
$$
Using Lemma A.1 we write 
$$
Op^W(f) =  Op^{AW}_\hbar(f) - \frac{\hbar}{4}  Op^{AW}_\hbar(\Delta f)
+ R(\hbar),
\eqno{(A.4)}
$$
where $|\langle R(\hbar)\rangle| \le C\hbar^2$.

Consider the matrix element of the first term on the right-hand side of
(A.4).  In Bargmann space this becomes
$$
\langle \psi_\alpha, Op^{AW}_\hbar(f) \psi_\alpha\rangle
= \frac{1}{\pi^l \hbar^{|\alpha|+l} \alpha!} \int \bar z^\alpha f(z,\hbar) 
z^\alpha
e^{-|z|^2/\hbar} \>dzd\bar z.
$$
Writing $f$ as a sum over $z^\gamma g_\gamma$ with $|\gamma|=4$, 
the estimate for a particular $\gamma$ is straightforward:
\begin{eqnarray*}
|\langle \psi_\alpha, Op^{AW}_\hbar(z^\gamma g_\gamma) \psi_\alpha\rangle|
&\le& M \frac{1}{\pi^l \hbar^{|\alpha|+l} \alpha!} \int 
\left|z^{\alpha}\right|^2 |z|^4
e^{-|z|^2/\hbar} \>dzd\bar z
\\
&=& M \hbar^{2} (|\alpha|+l)(|\alpha|+l+1).
\end{eqnarray*}

The second term on the right in (A.4) is handled in a similar way.
By assumption we can write $\Delta f = \sum_{|\eta|=2} z^\eta h_\eta(z,\hbar)$,
where $\sup|h_\eta| \le 12M$.  The estimate then proceeds exactly as above
(noting that there is an extra factor of $\hbar$ in front of this
term).\quadratino
\vskip 1.0cm\noindent

\end{document}